\documentclass[prb,superscriptaddress,floatfix,twocolumn]{revtex4-1}
\usepackage{graphicx,amssymb,bm,amsfonts,amsmath,graphics,color}
\usepackage[bookmarks=false,pdfstartview=FitH,colorlinks=true,citecolor=blue,linkcolor=blue]{hyperref}

\newcommand {\bisco}{Bi$_2$Sr$_2$CaCu$_2$O$_{8+\delta}$}

\newcommand {\KF}{$\bm{k}_F$}
\newcommand {\DI}{$\Delta$I}

\newcommand {\DIkF}{$\Delta$I$_{\bm{k}_F}$}

\begin{document}

\title{Nodal Quasiparticle Meltdown in Ultra-High Resolution Pump-Probe Angle-Resolved Photoemission}

\author{J. Graf}
\thanks{These authors contributed equally to this work.}
\affiliation{Materials Sciences Division, Lawrence Berkeley National Laboratory, Berkeley, CA 94720, USA}
\author {C. Jozwiak}
\thanks{These authors contributed equally to this work.}
\affiliation{Advanced Light Source, Lawrence Berkeley National Laboratory, Berkeley, CA 94720, USA}
\affiliation{Materials Sciences Division, Lawrence Berkeley National Laboratory, Berkeley, CA 94720, USA}
\author{C.L. Smallwood}
\affiliation{Department of Physics, University of California, Berkeley, CA 94720, USA} 
\author {H. Eisaki}
\affiliation {Superconducting Electronics Group, Electronics and Photonics Research Institute, National Institute of Advanced Industrial Science and Technology (AIST), 1-1-1 Central 2, Umezono, Tsukuba, Ibaraki 305-8568, Japan}
\author{R. A. Kaindl}
\affiliation{Materials Sciences Division, Lawrence Berkeley National Laboratory, Berkeley, CA 94720, USA}
\author{D.-H. Lee}
\affiliation{Department of Physics, University of California, Berkeley, CA 94720, USA}
\affiliation{Materials Sciences Division, Lawrence Berkeley National Laboratory, Berkeley, CA 94720, USA}
\author{A. Lanzara}\email{ALanzara@lbl.gov}
\affiliation{Department of Physics, University of California, Berkeley, CA 94720, USA}
\affiliation{Materials Sciences Division, Lawrence Berkeley National Laboratory, Berkeley, CA 94720, USA}


\maketitle

\textbf{
High-$T_c$ cuprate superconductors are characterized by a strong momentum-dependent anisotropy between the low energy excitations along the Brillouin zone diagonal (nodal direction) and those along the Brillouin zone face (antinodal direction).
Most obvious is the $d$-wave superconducting gap, with the largest magnitude found in the antinodal direction and no gap in the nodal direction.
Additionally, while antinodal quasiparticle excitations appear only below $T_c$, superconductivity is thought to be indifferent to nodal excitations as they are regarded robust and insensitive to $T_c$.
Here we reveal an unexpected tie between nodal quasiparticles and superconductivity using high resolution time- and angle-resolved photoemission on optimally doped \bisco. 
We observe a suppression of the nodal quasiparticle spectral weight following pump laser excitation and measure its recovery dynamics.
This suppression is dramatically enhanced in the superconducting state. 
These results reduce the nodal-antinodal dichotomy and challenge the conventional view of nodal excitation neutrality in superconductivity.
}

The electronic structures of high-$T_c$ cuprates are strongly momentum-dependent.
This is one reason why the momentum-resolved technique of angle-resolved photoemission spectroscopy (ARPES) has been a central tool in the field of high-temperature superconductivity.\cite{Damascelli03}
For example, coherent low energy excitations with momenta near the Brillouin zone face, or antinodal quasiparticles (QPs), are only observed below $T_c$ and have been linked to superfluid density.\cite{Feng00,Ding01}
They have therefore been the primary focus of ARPES studies.
In contrast, nodal QPs, with momenta along the Brillouin zone diagonal, have received less attention and are usually regarded as largely immune to the superconducting transition because they seem insensitive to perturbations such as disorder,\cite{Garg08,pan01,McElroy05} doping,\cite{Ando01,Zhou03,McElroy05} isotope exchange,\cite{Gweon04} charge ordering,\cite{Shen05,McElroy05,Vershinin04} and temperature.\cite{Valla99,Yusof02,Wei08,Kondo09}
Clearly, finding any strong dependencies of the nodal QPs will alter the conventional view and enrich our understanding of high temperature superconductivity.

Time resolution through pump-and-probe techniques adds a new dimension to ARPES by directly measuring how the electronic structure of a material responds to perturbations on femtosecond time scales.
Here we report a unique ultrafast time-resolved ARPES study of a high-$T_c$ cuprate superconductor.
Compared to previous time-resolved studies,\cite{Perfetti07,Schmitt08,Rettig10} the primary advantage of this work is an unprecedented momentum (angular) resolution ($\Delta k\sim$0.003 vs. 0.05~\AA$^{-1}$), on par with that of state-of-the-art ARPES.
This has allowed the time-resolved measurement of significantly sharper QP spectral peaks with strikingly larger peak-to-background ratios than previously reported.\cite{Perfetti07}
Additionally, a lower pump fluence is used ($\lesssim$40$\mu$J/cm$^2$ vs. $\sim$100$\mu$J/cm$^2$), which reduces pump-induced sample temperature increase and related thermal smearing of spectral features.
This allows us to uncover a surprising meltdown of nodal QP spectral weight following pump laser excitation.
This meltdown is only observed in the superconducting state and for QPs with binding energy less than the kink energy,\cite{Lanzara01} revealing a link between nodal QPs and superconductivity.

\section*{Nodal quasiparticle response to photoexcitation}

The experiment proceeds as follows.
A pump laser pulse (h$\nu$ = 1.5 eV) excites electrons from occupied to unoccupied states within 1.5 eV of the Fermi energy ($E_F$).
As shown by previous ultrafast optical studies, the photoexcited carriers relax to states near $E_F$ within 100 fs, generating a large non-equilibrium population of QPs and a depletion of the superconducting condensate.\cite{Han90, Stevens97, Demsar99, Kaindl00, Averitt01, Segre02, Gedik03, Gedik04, Kusar08, Liu08b}
The subsequent recombination of these non-equilibrium QPs is detected via photoemission with a probe laser pulse (h$\nu$ = 5.9 eV) as a function of a variable time delay between the pump and probe.
The laser pulse widths provide a time resolution $\approx$ 270 fs.

\begin{figure*}\centering\includegraphics[width=15.5cm]{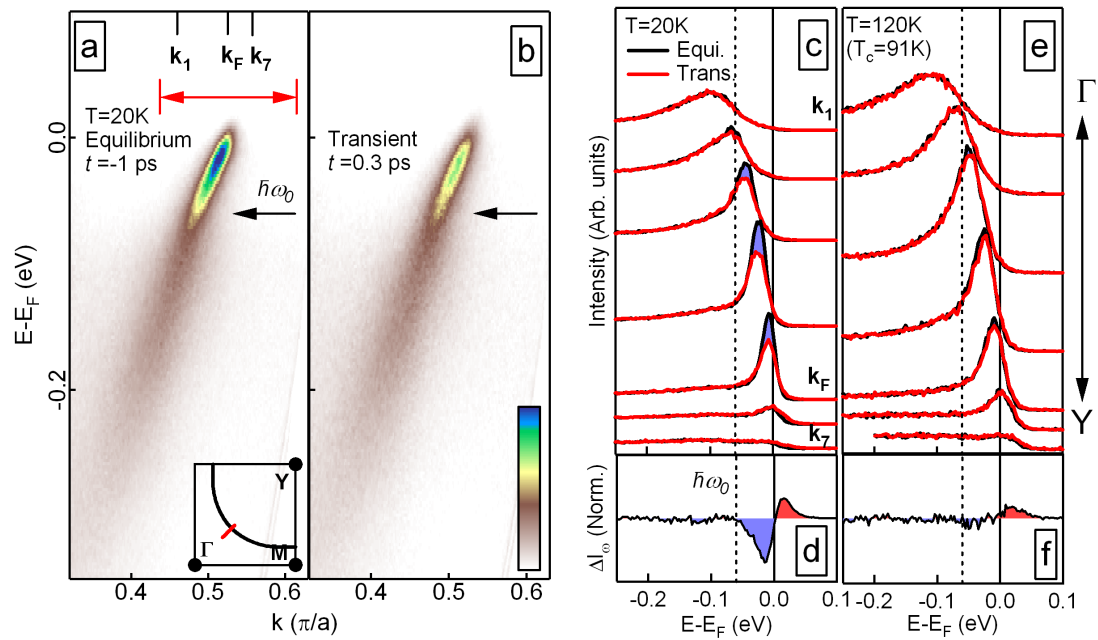}
\caption{\label{fig:no1}Nodal QP spectral weight suppression after an infrared pump pulse. 
(a) Energy-momentum map of equilibrium ARPES intensity before the pump pulse (negative delay time). The momentum location of the cut is shown on the Fermi surface inset. The well known dispersion kink is marked by the arrow.
(b) Corresponding map of the transient ARPES intensity 0.3 ps after the pump pulse.  The same color scale is used for both panels. 
(c) Energy distribution curves (EDCs) from $\bm{k_1}$ to $\bm{k_7}$ for the equilibrium (a, in black) and transient state (b, in red). Each EDC pair is separated by the same momentum value and vertically displaced by different amounts for clarity. 
Spectral weight loss (gain) is highlighted by the blue (red) areas.
(d) $\Delta I_{\omega}$, the difference between the transient and equilibrium EDCs, integrated through a momentum range centered around $k_F$ and shown by the red double arrow in (a).  
(e-f) Same comparison as in panel (c-d) but with the equilibrium sample temperature at 120 K ($T_c$ = 91 K).
(f) Plot is shown on the same scale as (d).}
\end{figure*}

Figure 1 compares the equilibrium electronic structure (before the arrival of the pump pulse) with the transient excited electronic structure at $t$ = 0.3 ps after excitation, above and below the superconducting transition.
For the low (high) temperature data, the bulk sample temperature was measured to be $T$ = 20 K (120 K) with a diode thermally connected to the crystal, and was found not to be measurably affected by the photoexcitation.
Low temperature ARPES intensity maps (energy vs. momentum) of the equilibrium and transient states are shown in panels (a) and (b), respectively, with identical color scales.
In both maps, one can identify a characteristic binding energy ($|E-E_F|$) $\hbar\omega_{0}\sim$ 60 meV (marked by arrows) that separates a low binding energy region between $E_F$ and $\hbar\omega_{0}$ characterized by a sharply defined dispersive peak, from a high binding energy region characterized by a poorly defined dispersive peak.
The crossover energy $\hbar\omega_{0}$ also marks the energy position of a kink in the nodal QP dispersion, which is a universal feature of the cuprates.\cite{Lanzara01}
An obvious pump-induced suppression of spectral intensity occurs which appears to be strictly confined to the low binding energy region (binding energy $<\hbar\omega_0$).

The data are compared more directly in panel (c), which shows raw energy distribution curves (EDCs - photoemission intensity as a function of energy at constant momentum) for the equilibrium (black) and transient (red) states, corresponding to the maps in panels (a) and (b), respectively.
The EDCs are normalized only by the total acquisition time and not by any additional high binding energy normalization scheme, and so a direct intensity comparison can accurately be made.
As in (a) and (b), the characteristic energy $\hbar\omega_{0}$ (marked by the vertical dotted line) separates sharp peaks from broad ones.
The sharp spectral peaks are signatures of coherent QPs.
The advantage of the high momentum resolution of the current work is clearly seen by the sharpness and intensity of the QP peaks.
Again, there is a clear suppression of spectral weight in the transient state, confined to the QPs at binding energies less than $\hbar\omega_{0}$.

Such suppression cannot be explained by thermal smearing of the Fermi edge due to the transient heating of the electronic temperature caused by the absorbed pump energy.
We find that the transient electrons around $E_F$ quickly thermalize and follow a Fermi-Dirac distribution within 100 fs of the ultrafast pump pulse, in close agreement with Ref.~\onlinecite{Perfetti07}.
Therefore the transient electronic temperature at each delay time can be obtained by direct measurement of the Fermi edge width of the corresponding transient ARPES signal (see Supplementary Information for details).
The electronic temperature of the transient data of Fig.~1 is thus found to be $<100$ K\@, and does not account for the observed spectral changes.
This is clear from Fig.~\ref{fig:no1}(c) by the significant loss of spectral weight at binding energies larger than 4$k_BT$ = 34 meV\@, which therefore cannot be attributed to Fermi edge thermal smearing.

\begin{figure*}\centering\includegraphics[width=15cm]{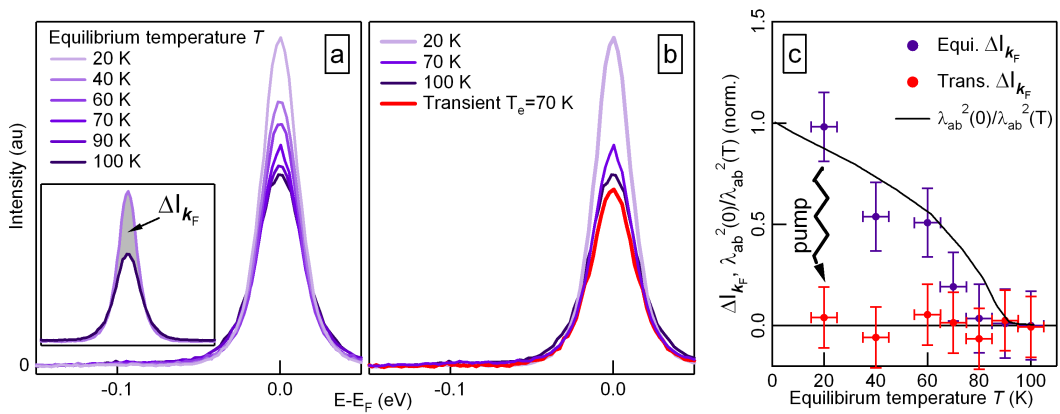}
\caption{\label{fig:no2}Comparison between equilibrium temperature driven and optical pump driven spectral weight suppression.
(a) Symmetrized equilibrium EDCs at \KF\ along the nodal direction at several equilibrium temperatures.  The inset defines \DIkF\ as the area difference between two EDCs.
(b) Selected equilibrium EDCs from (a) compared with a transient EDC with an equilibrium temperature of 20 K and a transient electronic temperature of 70 K\@.
(c) Equilibrium and transient \DIkF\ as a function of equilibrium temperature.
The superfluid density estimated from the penetration depth, $\lambda_{ab}$, measured by microwave spectroscopy [adapted from Ref.~\onlinecite{Jacobs95}] is shown for comparison.
The equilibrium and transient values of \DIkF\ are normalized by the equilibrium value of \DIkF\ at 20 K\@. The vertical error bars are based on the uncertainty on the determination of \KF, and the horizontal error bars are a generous estimation of error in the equilibrium temperature measured by silicon diode and a small temperature gradient between the diode and sample.
}\end{figure*}

Panel (d) shows the difference of the transient and equilibrium EDCs ($\Delta I_{\omega}$), integrated through a wide momentum range that captures the extent of observed spectral changes.
The negative blue (positive red) area under the curve illustrates the total nodal spectral weight lost (gained) after pumping as a function of binding energy.
Again, this clearly shows that photoinduced spectral change is confined to binding energies less than the kink energy, $\hbar\omega_{0}$.
Although it appears that the spectral loss (blue area) is larger than the spectral gain (red area), we note that such a comparison involves regions of vastly different signal intensities as the count rate below $E_F$ is far larger than that above.
Even small non-linearity effects common to the typical photoelectron detectors used\cite{Mannella2004a} may impact the relative magnitudes of the measured $\Delta I_{\omega}$ above and below $E_F$.

In panels (e) and (f), we present a similar comparison of EDCs associated with equilibrium and transient states, but this time with an equilibrium temperature of 120 K, above the superconducting transition $T_c$ = 91 K\@.
Like the low temperature data, $\hbar\omega_{0}$ separates sharp QP peaks from broad spectral features, and the QP peak sharpens as momentum $\bm{k}\rightarrow\bm{k}_F$ ($E\rightarrow E_F$).
The transient state also has an elevated electronic temperature, here determined to be $\sim$145 K\@ (see Supplementary Information).
In sharp contrast to panels (c,d), we do not observe a significant change in the transient state's QP spectral features in panels (e,f).
This shows that the surprising pump-induced loss of nodal QP spectral weight is dramatically sensitive to the superconducting state.

\section*{Nodal quasiparticle temperature dependence}

In Figure 2 we investigate the role of transient heating by the pump pulses in the nodal QP spectral weight suppression.
This was done by performing a standard equilibrium temperature dependence.
Panel (a) shows the resulting EDCs, symmetrized about $E_F$, at \KF\ for several temperatures from 20 K to above $T_c$.
The symmetrization removes thermal effects that enter through the Fermi-Dirac distribution by effectively canceling out the Fermi-Dirac function.\cite{Norman98}
Unlike previous synchrotron data,\cite{Yusof02,Wei08,Kondo09} our laser data show a substantial dependence of the nodal QP spectral weight on temperature.
This remarkable new observation, which is likely due to the enhanced bulk sensitivity achieved with relatively low photon energy by laser-ARPES,\cite{Casey08} on its own challenges the notion of complete robustness of nodal QPs.

\begin{figure*}\centering\includegraphics[width=16cm]{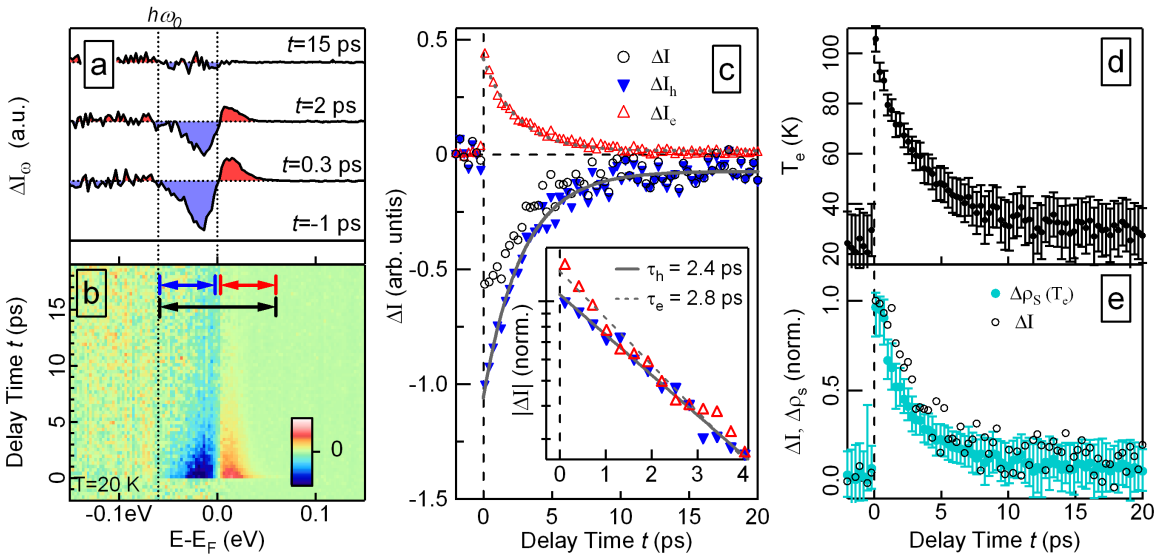}
\caption{\label{fig:no3}Time dependence of the pump-induced spectral changes for an equilibrium temperature of $T$ = 20 K\@.
(a) $\Delta I_{\omega}$, the difference of transient and equilibrium EDCs, integrated through the momentum range shown by the red double arrow in Fig.~\ref{fig:no1}, equivalent to Fig.~\ref{fig:no1}(d), at various delay times.  The curves are shifted vertically for clarity.
(b) Color map of $\Delta I_{\omega}$ as a function of delay times.  The curves in (a) are horizontal intensity profiles from (b).
(c) Angle and energy integrated spectral weight changes (\DI).  Open red triangles (filled blue triangles) correspond to the change in spectral intensity above (below) $E_F$, termed electrons (holes), and correspond to vertical intensity profiles of $\Delta I_{\omega}$ in (b), integrating along the energy range spanned by the red (blue) double arrow.  The solid and dashed gray lines are exponential fits.  The open black circles correspond to the sum of the red and blue triangles, or integrating $\Delta I_{\omega}$ along the energy range spanned by the black double arrow.  The inset compares the $\Delta I_{e}$ and $\Delta I_{h}$ dynamics in a smaller time range, with $\Delta I_{e}$ inverted and normalized.
(d) Electronic temperature $T_e$ determined from the width of the Fermi-Dirac distribution as a function of delay time. The error bars are from the fit standard deviation and the uncertainty in the experimental resolution and the space charging contribution\cite{Graf10}.
(e) \DI~(black circles from (c)) compared with the estimated minimal change of superfluid density ($\Delta\rho_s(T_e)$) (see text) as a function of delay time (light blue markers).
}\end{figure*}

In panel (b), we directly compare the temperature dependence of the equilibrium QP spectral weight with pump-induced spectral weight suppression.
Here, the electronic temperature of the pumped transient state is 70$\pm$5 K.
As described above, the transient electronic temperature is obtained by the direct measurement of the Fermi edge width of the transient ARPES signal (see Supplementary Information for details).
The lower transient electronic temperature relative to that in Fig.~1 was achieved by reducing the fluence of the pump pulse from $\sim$40 to $\sim$25 $\mu$J/cm$^2$.
Although the electronic temperature of the transient EDC is only $\sim$70 K\@, it exhibits a similar or greater loss of spectral weight than the 100 K equilibrium EDC.
This suggests that optical pumping induces a measurable effect beyond what is consistent with merely increasing the sample temperature.
This suggests that the transient nodal QP spectral weight does not simply follow the transient electronic temperature, a concept which is consistent with the fact that the transient state is not at equilibrium.
However, it should be noted that the inherent difficulties of precise comparisons of equilibrium temperature dependent data (e.g. maintaining the same sample spot and momentum-space position) place significant error bars on such direct comparisons (see Fig.~\ref{fig:no3}c).

In panel (c) we compare the equilibrium bulk temperature dependence of the QP spectral weight suppression in the equilibrium and transient states. 
Here we define the equilibrium (transient) \DIkF as the difference between the areas under the equilibrium (transient) EDC at $T$ and the equilibrium (transient) normal state EDC ($T$ = 100 K). (See illustration in the inset of Fig.~2(a).)
We make the following two observations: (1) the equilibrium value of \DIkF\ (purple markers) has a temperature dependence which is remarkably similar to the antinodal QP spectral weight \cite{Feng00,Ding01} and to the superfluid density, shown for comparison in the same figure (solid black line); and (2) the transient state \DIkF\ (red markers) is temperature independent.
Point (1) suggests that the equilibrium value of \DIkF\ is proportional to the superfluid density.
This finding is quite surprising as it contradicts the conventional view that the nodal QPs are uninvolved with and insensitive to superconductivity, and argues that the dichotomy between nodal and antinodal QPs is not as extensive as often thought.
However, despite the similarity between the temperature dependence of the nodal and antinodal QP spectral weight, there remain two important distinctions.  
First, the antinodal QP peak is almost entirely suppressed above $T_c$ while a finite component of the nodal QP peak is robust and persists above $T_c$. 
Second, the antinodal QP peak, despite its strong temperature dependence, is believed to satisfy a sum rule\cite{Randeria95} (the total spectral weight at a given momentum is conserved). 
In contrast, our data show that the nodal QP peak spectral weight is not conserved within our experimentally accessible energy window, either with increasing equilibrium temperature or with photoexcitation.
It is possible that the missing spectral weight is redistributed to higher energies and/or different momenta\cite{Basov99,Molegraaf02,Phillips04}, for instance to the antinodal region.

These findings suggest that the transient state superfluid density is nearly zero regardless of the equilibrium temperature.
Indeed it is well established by ultrafast THz experiments that an optical pump pulse can induce a suppression of superfluid density in the transient state,\cite{Averitt01,Kaindl05} and the current saturation effect is consistent with other optical pump-probe experiments that have found the depletion of the superconducting condensate is nearly complete at pump fluences below what is used here.\cite{Segre02,Carnahan04}
In addition, the fact that the transient state nodal spectral weight is very similar to the equilibrium nodal spectral weight in the normal state also supports the conclusion that the transient state superfluid density has been suppressed to zero by the optical pump while the electronic temperature is below $T_c$.
Again, we stress that the transient state is not in equilibrium, and that these observations may reflect the possibility that the electronic temperature cools at a faster rate than the superconducting condensate reforms.
Such an effect may be expected in this non-equilibrium state since the rate of condensate formation is additionally constrained by energy and momentum conservation of the quasiparticle interactions underlying the charge pairing.

\section*{Dynamics of nodal quasiparticle response}

In Figure 3 we report the recovery of the nodal QP spectral weight as a function of delay time, determined from the angle-integrated photoemission intensity for an equilibrium temperature of 20 K\@.
To focus on spectral change, we show the spectral intensity with the equilibrium data ($t$ = -1 ps) subtracted.
Similar to Figs.~\ref{fig:no1}(d,f), panel (a) shows the angle integrated spectral intensity change between transient and equilibrium EDCs, $\Delta I_{\omega}$, at various delay times, with the curve at $t=-1$ ps being exactly flat by definition.
Panel (b) shows a false-color map of the spectral intensity change, built up by $\Delta I_{\omega}$ curves at many delay times.
Two main features appear in both panels: a pump-induced increase in spectral intensity above $E_F$ (red feature) that can be partially understood as the thermally excited population of the states above $E_F$, and a spectral weight suppression confined between $E_F$ and $\hbar\omega_{0}$ (blue feature).

The dynamics of these spectral changes, $\Delta I_e$ and $\Delta I_h$, are shown in panel (c) integrated through the energy ranges marked by the red and blue arrows, respectively, in (b).
The total change in spectral weight, \DI, is integrated through the energy range marked by the black arrow.
The dashed and solid gray lines are single exponential least-square fits to $\Delta I_e$ and $\Delta I_h$ with time constants of 2.4 and 2.8 ps, respectively.
These values agree well with those observed at similar pump fluences in optical reflectivity measurements of underdoped \bisco.\cite{Liu08b}
The two dynamics are shown renormalized and on a log scale in the inset to provide a more direct comparison, and a difference in dynamics is observed.
Indeed, $\Delta I_e$ appears to have a kink, indicating an additional, fast component which is not apparent in $\Delta I_h$, although improved statistics are required to more accurately determine this aspect in future works.
In a simple approach, one may expect that the dynamics are governed by multiple components, one of which is purely controlled by the electronic temperature through the smearing of the Fermi edge and acts symmetrically above and below $E_F$, while other components may reflect dynamics which do not trivially follow the electronic temperature.

In panel (d) we show the time dependence of the transient electronic temperature, $T_{e}$, determined by fitting the Fermi edge of the angle-integrated data at the node with a Fermi-Dirac distribution function.
The time dependence shows that the equilibrium temperature of 20 K is recovered before the following pump pulse, $\sim1  \mu$s later.
This guarantees that there is minimal residual heating from pulse to pulse in all our experiments.
When these transient electronic temperature data are combined with microwave spectroscopy data of the superfluid density\cite{Jacobs95} shown in Fig.~2(c), a minimum value for the change in the superfluid density ($\Delta\rho_s (T_e)$), based solely on the electronic temperature, is obtained and shown in light blue in Fig.~\ref{fig:no3} (e).
This is directly compared with the normalized absolute value of the net \DI\ from panel (c).
The similarity between the net \DI\ and $\Delta\rho_s (T_e)$ in this panel is consistent with our proposal that the lost nodal spectral weight is proportional to the superfluid density.
More precisely, the fact that \DI\ does not decay on a faster time scale than $\Delta\rho_s (T_e)$ is also consistent with this proposal.
That the data possibly show that the dynamics of \DI\ lag behind $\Delta\rho_s (T_e)$ may also be consistent with the notion that the time scale for Cooper pair formation in \bisco \cite{Kaindl00} occurs at slower time scales than the present decay of the electronic temperature, thus limiting the decay of \DI\ to be slower than that of $\Delta\rho_s (T_e)$.
However, better statistics in future works are necessary for a more rigorous comparison to establish the exact dynamic of \DI.

In summary, we propose that the observed temperature- and/or pump- driven suppression of QP spectral weight is due to an increase in superconducting Cooper pair phase fluctuations\cite{Emery95,Corson99,Khodas10} corresponding to a loss in superfluid density.
Further work with this approach is motivated to expand the study to other regions of momentum space as one may expect the behavior to be different towards the antinode.

\section*{Methods}
Pump and probe photon beams were generated from a cavity-dumped, mode-locked Ti:Sapphire oscillator (Coherent Mira) providing $\sim$ 150 fs pulses of 840 nm at $\lesssim$ 1 MHz repetition rate.
A fraction of the beam was variably delayed and focused onto the sample as the pump with a spot size of 100 $\mu$m diameter and fluences up to 40 $\mu$J/cm$^2$.
The remaining beam was frequency quadrupled to form the probe beam with $h\nu=5.9$ eV and a measured bandwidth of 15 meV and focused onto the same sample spot with a size of $\sim$ 40 $\times$ 60 $\mu$m.
The combined beams were cross-correlated in time with a FWHM of 0.27 ps.

ARPES measurements were taken with a 150 mm hemispherical analyzer with 2D imaging detector (SPECS Phoibos 150).
The total experimental energy resolution, including probe beam bandwidth, was measured to be 21 meV.
The angular resolution of $<0.3^{\circ}$ along the nodal direction corresponds to a momentum resolution of $<0.003$ \AA$^{-1}$ at the given photoelectron kinetic energies and emission angles.
Probe and pump photons were s-polarized, with the polarization vector perpendicular to the direction of the nodal momentum space cut measured.

Samples were mounted on a 6-axis manipulator, cleaved at $4\times10^{-11}$ torr, and directly oriented by taking ARPES Fermi surface maps.  The data for Figs.~1(a-d) and 3 are from samples cleaved and maintained at 20 K\@, the data for Figs.~1(e-f) are from samples cleaved and maintained at 120 K\@, and the data for Fig. 2 are from a sample cleaved at 100 K\@, with the T dependence taken at sequentially lower temperatures.

\section*{Correspondence}
Correspondence and requests for materials should be addressed to A.L.

\begin{acknowledgments}
 We thank Z. Hussain for support in the initial stage of the project and J. Orenstein and W. Zhang for useful discussions. This work was supported by the Director, Office of Science, Office of Basic Energy Sciences, Materials Sciences and Engineering Division, of the U.S. Department of Energy under Contract No. DE-AC02-05CH11231.
 \end{acknowledgments}

\section*{Author Contributions}
J.G. and C.J. designed and built the laser-ARPES system.  R.A.K. contributed to the design concept of the laser-ARPES system.  J.G. carried out the experiment.  J.G., C.J., and C.S. are responsible for data analysis.  H.E. prepared the samples.  A.L. is responsible for the experimental concept, planning, and infrastructure.  All authors contributed to the interpretation and manuscript.
\section*{Additional information}
The authors declare that they have no competing financial interests.
Supplementary Information accompanies this paper.


\begin{thebibliography}{10}
\expandafter\ifx\csname url\endcsname\relax
  \def\url#1{\texttt{#1}}\fi
\expandafter\ifx\csname urlprefix\endcsname\relax\def\urlprefix{URL }\fi
\providecommand{\bibinfo}[2]{#2}
\providecommand{\eprint}[2][]{\url{#2}}

\bibitem{Damascelli03}
\bibinfo{author}{Damascelli, A.}, \bibinfo{author}{Hussain, Z.} \&
  \bibinfo{author}{Shen, Z.-X.}
\newblock \bibinfo{title}{Angle-resolved photoemission studies of the cuprate
  superconductors}.
\newblock \emph{\bibinfo{journal}{Rev. of Mod. Phys.}}
  \textbf{\bibinfo{volume}{75}}, \bibinfo{pages}{473--541} (\bibinfo{year}{2003}).

\bibitem{Feng00}
\bibinfo{author}{Feng, D.~L.} \emph{et~al.}
\newblock \bibinfo{title}{Signature of superfluid density in the
  single-particle excitation spectrum of
  Bi$_{2}$Sr$_{2}$CaCu$_{2}$O$_{8+\delta}$}.
\newblock \emph{\bibinfo{journal}{Science}} \textbf{\bibinfo{volume}{289}},
  \bibinfo{pages}{277--281} (\bibinfo{year}{2000}).

\bibitem{Ding01}
\bibinfo{author}{Ding, H.} \emph{et~al.}
\newblock \bibinfo{title}{Coherent quasiparticle weight and its connection to
  high- $T_c$ superconductivity from angle-resolved photoemission}.
\newblock \emph{\bibinfo{journal}{Phys. Rev. Lett.}}
  \textbf{\bibinfo{volume}{87}}, \bibinfo{pages}{227001}
  (\bibinfo{year}{2001}).

\bibitem{Garg08}
\bibinfo{author}{Garg, A.}, \bibinfo{author}{Randeria, M.} \&
  \bibinfo{author}{Trivedi, N.}
\newblock \bibinfo{title}{Strong correlations make high-temperature
  superconductors robust against disorder}.
\newblock \emph{\bibinfo{journal}{Nat Phys}} \textbf{\bibinfo{volume}{4}},
  \bibinfo{pages}{762--765} (\bibinfo{year}{2008}).

\bibitem{pan01}
\bibinfo{author}{Pan, S.~H.} \emph{et~al.}
\newblock \bibinfo{title}{Microscopic electronic inhomogeneity in the high-tc
  superconductor Bi$_{2}$Sr$_{2}$CaCu$_{2}$O$_{8+x}$}.
\newblock \emph{\bibinfo{journal}{Nature}} \textbf{\bibinfo{volume}{413}},
  \bibinfo{pages}{282--285} (\bibinfo{year}{2001}).

\bibitem{McElroy05}
\bibinfo{author}{McElroy, K.} \emph{et~al.}
\newblock \bibinfo{title}{Coincidence of {C}heckerboard {C}harge {O}rder and
  {A}ntinodal {S}tate {D}ecoherence in {S}trongly {U}nderdoped
  {S}uperconducting {B}i$_2${S}r$_2${C}a{C}u$_2${O}$_{8+\delta}$}.
\newblock \emph{\bibinfo{journal}{Phys. Rev. Lett.}}
  \textbf{\bibinfo{volume}{94}}, \bibinfo{pages}{197005}
  (\bibinfo{year}{2005}).

\bibitem{Ando01}
\bibinfo{author}{Ando, Y.}, \bibinfo{author}{Lavrov, A.~N.},
  \bibinfo{author}{Komiya, S.}, \bibinfo{author}{Segawa, K.} \&
  \bibinfo{author}{Sun, X.~F.}
\newblock \bibinfo{title}{Mobility of the doped holes and the antiferromagnetic
  correlations in underdoped high- $T_c$ cuprates}.
\newblock \emph{\bibinfo{journal}{Phys. Rev. Lett.}}
  \textbf{\bibinfo{volume}{87}}, \bibinfo{pages}{017001}
  (\bibinfo{year}{2001}).

\bibitem{Zhou03}
\bibinfo{author}{Zhou, X.~J.} \emph{et~al.}
\newblock \bibinfo{title}{High-temperature superconductors: {U}niversal nodal
  {F}ermi velocity}.
\newblock \emph{\bibinfo{journal}{Nature}} \textbf{\bibinfo{volume}{423}},
  \bibinfo{pages}{398} (\bibinfo{year}{2003}).

\bibitem{Gweon04}
\bibinfo{author}{Gweon, G.-H.} \emph{et~al.}
\newblock \bibinfo{title}{An unusual isotope effect in a
  high-transition-temperature superconductor}.
\newblock \emph{\bibinfo{journal}{Nature}} \textbf{\bibinfo{volume}{430}},
  \bibinfo{pages}{187--190} (\bibinfo{year}{2004}).

\bibitem{Shen05}
\bibinfo{author}{Shen, K.~M.} \emph{et~al.}
\newblock \bibinfo{title}{Nodal {Q}uasiparticles and {A}ntinodal {C}harge
  {O}rdering in {C}a$_{2-x}${N}a$_{x}${C}u{O}$_{2}${C}l$_{2}$}.
\newblock \emph{\bibinfo{journal}{Science}} \textbf{\bibinfo{volume}{307}},
  \bibinfo{pages}{901--904} (\bibinfo{year}{2005}).

\bibitem{Vershinin04}
\bibinfo{author}{Vershinin, M.} \emph{et~al.}
\newblock \bibinfo{title}{Local ordering in the pseudogap state of the high-tc
  superconductor Bi$_{2}$Sr$_{2}$CaCu$_{2}$O$_{8+\delta}$}.
\newblock \emph{\bibinfo{journal}{Science}} \textbf{\bibinfo{volume}{303}},
  \bibinfo{pages}{1995--1998} (\bibinfo{year}{2004}).

\bibitem{Valla99}
\bibinfo{author}{Valla, T.} \emph{et~al.}
\newblock \bibinfo{title}{Evidence for quantum critical behavior in the
  optimally doped cuprate Bi$_{2}$Sr$_{2}$CaCu$_{2}$O$_{8+\delta}$}.
\newblock \emph{\bibinfo{journal}{Science}} \textbf{\bibinfo{volume}{285}},
  \bibinfo{pages}{2110--2113} (\bibinfo{year}{1999}).

\bibitem{Yusof02}
\bibinfo{author}{Yusof, Z.~M.} \emph{et~al.}
\newblock \bibinfo{title}{Quasiparticle liquid in the highly overdoped
  Bi$_{2}$Sr$_{2}$CaCu$_{2}$O$_{8+\delta }$}.
\newblock \emph{\bibinfo{journal}{Phys. Rev. Lett.}}
  \textbf{\bibinfo{volume}{88}}, \bibinfo{pages}{167006}
  (\bibinfo{year}{2002}).

\bibitem{Wei08}
\bibinfo{author}{Wei, J.} \emph{et~al.}
\newblock \bibinfo{title}{Superconducting coherence peak in the electronic
  excitations of a single-layer Bi$_{2}$Sr$_{1.6}$La$_{0.4}$CuO$_{6+\delta}$
  cuprate superconductor}.
\newblock \emph{\bibinfo{journal}{Phys. Rev. Lett.}}
  \textbf{\bibinfo{volume}{101}}, \bibinfo{pages}{097005}
  (\bibinfo{year}{2008}).

\bibitem{Kondo09}
\bibinfo{author}{Kondo, T.}, \bibinfo{author}{Khasanov, R.},
  \bibinfo{author}{Takeuchi, T.}, \bibinfo{author}{Schmalian, J.} \&
  \bibinfo{author}{Kaminski, A.}
\newblock \bibinfo{title}{Competition between the pseudogap and
  superconductivity in the high-$T_c$ copper oxides}.
\newblock \emph{\bibinfo{journal}{Nature}} \textbf{\bibinfo{volume}{457}},
  \bibinfo{pages}{296--300} (\bibinfo{year}{2009}).

\bibitem{Perfetti07}
\bibinfo{author}{Perfetti, L.} \emph{et~al.}
\newblock \bibinfo{title}{Ultrafast electron relaxation in superconducting
  Bi$_{2}$Sr$_{2}$CaCu$_{2}$O$_{8 + \delta}$ by time-resolved photoelectron
  spectroscopy}.
\newblock \emph{\bibinfo{journal}{Phys. Rev. Lett.}}
  \textbf{\bibinfo{volume}{99}}, \bibinfo{pages}{197001}
  (\bibinfo{year}{2007}).

\bibitem{Schmitt08}
\bibinfo{author}{Schmitt, F.} \emph{et~al.}
\newblock \bibinfo{title}{Transient electronic structure and melting of a
  charge density wave in TbTe$_{3}$}.
\newblock \emph{\bibinfo{journal}{Science}} \textbf{\bibinfo{volume}{321}},
  \bibinfo{pages}{1649--1652} (\bibinfo{year}{2008}).

\bibitem{Rettig10}
\bibinfo{author}{Rettig, L.} \emph{et~al.}
\newblock \bibinfo{title}{Electron-phonon coupling and momentum dependent electron dynamics in EuFe$_2$As$_2$ using time- and angle-resolved photoemission spectroscopy}.
\newblock {Preprint at http://arxiv.org/abs/1008.1561v2} (\bibinfo{year}{2010}).

\bibitem{Lanzara01}
\bibinfo{author}{Lanzara, A.} \emph{et~al.}
\newblock \bibinfo{title}{Evidence for ubiquitous strong electron-phonon
  coupling in high-temperature superconductors}.
\newblock \emph{\bibinfo{journal}{Nature}} \textbf{\bibinfo{volume}{412}},
  \bibinfo{pages}{510--514} (\bibinfo{year}{2001}).

\bibitem{Han90}
\bibinfo{author}{Han, S.~G.}, \bibinfo{author}{Vardeny, Z.~V.},
  \bibinfo{author}{Wong, K.~S.}, \bibinfo{author}{Symko, O.~G.} \&
  \bibinfo{author}{Koren, G.}
\newblock \bibinfo{title}{Femtosecond optical detection of quasiparticle
  dynamics in high-$T_c$ YBa$_{2}$Cu$_{3}$O$_{7-\delta}$ superconducting thin
  films}.
\newblock \emph{\bibinfo{journal}{Phys. Rev. Lett.}}
  \textbf{\bibinfo{volume}{65}}, \bibinfo{pages}{2708--2711} (\bibinfo{year}{1990}).

\bibitem{Stevens97}
\bibinfo{author}{Stevens, C.~J.} \emph{et~al.}
\newblock \bibinfo{title}{Evidence for two-component high-temperature
  superconductivity in the femtosecond optical response of
  YBa$_{2}$Cu$_{3}$O$_{7-\delta}$}.
\newblock \emph{\bibinfo{journal}{Phys. Rev. Lett.}}
  \textbf{\bibinfo{volume}{78}}, \bibinfo{pages}{2212--2215} (\bibinfo{year}{1997}).

\bibitem{Demsar99}
\bibinfo{author}{Demsar, J.}, \bibinfo{author}{Podobnik, B.},
  \bibinfo{author}{Kabanov, V.~V.}, \bibinfo{author}{Wolf, T.} \&
  \bibinfo{author}{Mihailovic, D.}
\newblock \bibinfo{title}{Superconducting gap {$\Delta _{c}$}, the pseudogap
  {$\Delta _{p}$}, and pair fluctuations above $T_c$ in overdoped
  Y$_{1-x}$Ca$_{x}$Ba$_{2}$Cu$_{3}$O$_{7- \delta}$ from femtosecond time-domain
  spectroscopy}.
\newblock \emph{\bibinfo{journal}{Phys. Rev. Lett.}}
  \textbf{\bibinfo{volume}{82}}, \bibinfo{pages}{4918--4921}
  (\bibinfo{year}{1999}).

\bibitem{Kaindl00}
\bibinfo{author}{Kaindl, R.~A.} \emph{et~al.}
\newblock \bibinfo{title}{Ultrafast mid-infrared response of
  YBa$_{2}$Cu$_{3}$O$_{7-\delta}$}.
\newblock \emph{\bibinfo{journal}{Science}} \textbf{\bibinfo{volume}{287}},
  \bibinfo{pages}{470--473} (\bibinfo{year}{2000}).

\bibitem{Averitt01}
\bibinfo{author}{Averitt, R.~D.} \emph{et~al.}
\newblock \bibinfo{title}{Nonequilibrium superconductivity and quasiparticle
  dynamics in YBa$_{2}$Cu$_{3}$O$_{7-\delta}$}.
\newblock \emph{\bibinfo{journal}{Phys. Rev. B}} \textbf{\bibinfo{volume}{63}},
  \bibinfo{pages}{140502(R)} (\bibinfo{year}{2001}).

\bibitem{Segre02}
\bibinfo{author}{Segre, G.~P.} \emph{et~al.}
\newblock \bibinfo{title}{Photoinduced changes of reflectivity in single
  crystals of YBa$_{2}$Cu$_{3}$O$_{6.5}$ (ortho ii)}.
\newblock \emph{\bibinfo{journal}{Phys. Rev. Lett.}}
  \textbf{\bibinfo{volume}{88}}, \bibinfo{pages}{137001}
  (\bibinfo{year}{2002}).

\bibitem{Gedik03}
\bibinfo{author}{Gedik, N.}, \bibinfo{author}{Orenstein, J.},
  \bibinfo{author}{Liang, R.}, \bibinfo{author}{Bonn, D.~A.} \&
  \bibinfo{author}{Hardy, W.~N.}
\newblock \bibinfo{title}{Diffusion of nonequilibrium quasi-particles in a
  cuprate superconductor}.
\newblock \emph{\bibinfo{journal}{Science}} \textbf{\bibinfo{volume}{300}},
  \bibinfo{pages}{1410--1412} (\bibinfo{year}{2003}).

\bibitem{Gedik04}
\bibinfo{author}{Gedik, N.} \emph{et~al.}
\newblock \bibinfo{title}{Single-quasiparticle stability and quasiparticle-pair
  decay in YBa$_{2}$Cu$_{3}$O$_{6.5}$}.
\newblock \emph{\bibinfo{journal}{Phys. Rev. B}} \textbf{\bibinfo{volume}{70}},
  \bibinfo{pages}{014504} (\bibinfo{year}{2004}).

\bibitem{Kusar08}
\bibinfo{author}{Kusar, P.} \emph{et~al.}
\newblock \bibinfo{title}{Controlled vaporization of the superconducting
  condensate in cuprate superconductors by femtosecond photoexcitation}.
\newblock \emph{\bibinfo{journal}{Phys. Rev. Lett.}}
  \textbf{\bibinfo{volume}{101}}, \bibinfo{pages}{227001}
  (\bibinfo{year}{2008}).

\bibitem{Liu08b}
\bibinfo{author}{Liu, Y.~H.} \emph{et~al.}
\newblock \bibinfo{title}{Direct observation of the coexistence of the
  pseudogap and superconducting quasiparticles in
  Bi$_{2}$Sr$_{2}$CaCu$_{2}$O$_{8+y}$ by time-resolved optical spectroscopy}.
\newblock \emph{\bibinfo{journal}{Phys. Rev. Lett.}}
  \textbf{\bibinfo{volume}{101}}, \bibinfo{pages}{137003}
  (\bibinfo{year}{2008}).

\bibitem{Mannella2004a}
\bibinfo{author}{Mannella, N.} \emph{et~al.}
\newblock \bibinfo{title}{Correction of non-linearity effects in detectors for
  electron spectroscopy}.
\newblock \emph{\bibinfo{journal}{J. Electron Spectrosc. Relat. Phenom.}} \textbf{\bibinfo{volume}{141}}, \bibinfo{pages}{45--59}
  (\bibinfo{year}{2004}).

\bibitem{Jacobs95}
\bibinfo{author}{Jacobs, T.}, \bibinfo{author}{Sridhar, S.},
  \bibinfo{author}{Li, Q.}, \bibinfo{author}{Gu, G.~D.} \&
  \bibinfo{author}{Koshizuka, N.}
\newblock \bibinfo{title}{In-plane and c-axis microwave penetration depth of
  Bi$_{2}$Sr$_{2}$CaCu$_{2}$O$_{8+\delta}$ crystals}.
\newblock \emph{\bibinfo{journal}{Phys. Rev. Lett.}}
  \textbf{\bibinfo{volume}{75}}, \bibinfo{pages}{4516--4519}
  (\bibinfo{year}{1995}).

\bibitem{Norman98}
\bibinfo{author}{Norman, M.~R.} \emph{et~al.}
\newblock \bibinfo{title}{Destruction of the fermi surface in underdoped
  high-$T_c$ superconductors}.
\newblock \emph{\bibinfo{journal}{Nature}} \textbf{\bibinfo{volume}{392}},
  \bibinfo{pages}{157--160} (\bibinfo{year}{1998}).

\bibitem{Casey08}
\bibinfo{author}{Casey, P.~A.}, \bibinfo{author}{Koralek, J.~D.},
  \bibinfo{author}{Plumb, N.~C.}, \bibinfo{author}{Dessau, D.~S.} \&
  \bibinfo{author}{Anderson, P.~W.}
\newblock \bibinfo{title}{Accurate theoretical fits to laser-excited
  photoemission spectra in the normal phase of high-temperature
  superconductors}.
\newblock \emph{\bibinfo{journal}{Nat. Phys.}} \textbf{\bibinfo{volume}{4}},
  \bibinfo{pages}{210--212} (\bibinfo{year}{2008}).

\bibitem{Randeria95}
\bibinfo{author}{Randeria, M.} \emph{et~al.}
\newblock \bibinfo{title}{Momentum distribution sum rule for angle-resolved
  photoemission}.
\newblock \emph{\bibinfo{journal}{Phys. Rev. Lett.}}
  \textbf{\bibinfo{volume}{74}}, \bibinfo{pages}{4951--4954} (\bibinfo{year}{1995}).

\bibitem{Basov99}
\bibinfo{author}{{Basov}, D.~N.} \emph{et~al.}
\newblock \bibinfo{title}{Sum rules and interlayer conductivity of high-$T_c$
  cuprates}.
\newblock \emph{\bibinfo{journal}{Science}} \textbf{\bibinfo{volume}{283}},
  \bibinfo{pages}{49--52} (\bibinfo{year}{1999}).

\bibitem{Molegraaf02}
\bibinfo{author}{Molegraaf, H. J.~A.}, \bibinfo{author}{Presura, C.},
  \bibinfo{author}{van~der Marel, D.}, \bibinfo{author}{Kes, P.~H.} \&
  \bibinfo{author}{Li, M.}
\newblock \bibinfo{title}{Superconductivity-induced transfer of in-plane
  spectral weight in Bi$_{2}$Sr$_{2}$CaCu$_{2}$O$_{8+\delta}$}.
\newblock \emph{\bibinfo{journal}{Science}} \textbf{\bibinfo{volume}{295}},
  \bibinfo{pages}{2239--2241} (\bibinfo{year}{2002}).

\bibitem{Phillips04}
\bibinfo{author}{{Phillips}, P.}, \bibinfo{author}{{Galanakis}, D.} \&
  \bibinfo{author}{{Stanescu}, T.~D.}
\newblock \bibinfo{title}{Absence of asymptotic freedom in doped Mott
  insulators: Breakdown of strong coupling expansions}.
\newblock \emph{\bibinfo{journal}{Phys. Rev. Lett.}}
  \textbf{\bibinfo{volume}{93}}, \bibinfo{pages}{267004}
  (\bibinfo{year}{2004}).

\bibitem{Kaindl05}
\bibinfo{author}{Kaindl, R.~A.}, \bibinfo{author}{Carnahan, M.~A.},
  \bibinfo{author}{Chemla, D.~S.}, \bibinfo{author}{Oh, S.} \&
  \bibinfo{author}{Eckstein, J.~N.}
\newblock \bibinfo{title}{Dynamics of cooper pair formation in
  Bi$_2$Sr$_2$CaCu$_2$O$_{8+\delta}$}.
\newblock \emph{\bibinfo{journal}{Phys. Rev. B}} \textbf{\bibinfo{volume}{72}},
  \bibinfo{pages}{060510(R)} (\bibinfo{year}{2005}).

\bibitem{Carnahan04}
\bibinfo{author}{Carnahan, M.~A.} \emph{et~al.}
\newblock \bibinfo{title}{Nonequilibrium thz conductivity of
  Bi$_{2}$Sr$_{2}$CaCu$_{2}$O$_{8+\delta}$}.
\newblock \emph{\bibinfo{journal}{Phys. C}} \textbf{\bibinfo{volume}{408-410}},
  \bibinfo{pages}{729--730} (\bibinfo{year}{2004}).

\bibitem{Graf10}
\bibinfo{author}{Graf, J.} \emph{et~al.}
\newblock \bibinfo{title}{Vacuum space charge effect in laser-based solid-state
  photoemission spectroscopy}.
\newblock \emph{\bibinfo{journal}{J. Appl. Phys.}}
  \textbf{\bibinfo{volume}{107}}, \bibinfo{pages}{014912}
  (\bibinfo{year}{2010}).

\bibitem{Emery95}
\bibinfo{author}{Emery, V.~J.} \& \bibinfo{author}{Kivelson, S.~A.}
\newblock \bibinfo{title}{Importance of phase fluctuations in superconductors
  with small superfluid density}.
\newblock \emph{\bibinfo{journal}{Nature}} \textbf{\bibinfo{volume}{374}},
  \bibinfo{pages}{434--437} (\bibinfo{year}{1995}).

\bibitem{Corson99}
\bibinfo{author}{Corson, J.}, \bibinfo{author}{Mallozzi, R.},
  \bibinfo{author}{Orenstein, J.}, \bibinfo{author}{Eckstein, J.~N.} \&
  \bibinfo{author}{Bozovic, I.}
\newblock \bibinfo{title}{Vanishing of phase coherence in underdoped
  Bi$_2$Sr$_2$CaCu$_2$O$_{8+\delta}$}.
\newblock \emph{\bibinfo{journal}{Nature}} \textbf{\bibinfo{volume}{398}},
  \bibinfo{pages}{221--223} (\bibinfo{year}{1999}).

\bibitem{Khodas10}
\bibinfo{author}{Khodas, M.} \& \bibinfo{author}{Tsvelik, A.~M.}
\newblock \bibinfo{title}{Influence of thermal phase fluctuations on the
  spectral function for a two-dimensional d-wave superconductor}.
\newblock \emph{\bibinfo{journal}{Phys. Rev. B}} \textbf{\bibinfo{volume}{81}},
  \bibinfo{pages}{094514} (\bibinfo{year}{2010}).

\end{thebibliography}
\end{document}